\documentstyle[psfig,referee]{mn}

\input{epsf}

\begin{document}

\title[An Early-Time Infrared and Optical Study of the Type~Ia
Supernova~1998bu in M~96] {An Early-Time Infrared and Optical Study
of the Type~Ia Supernova~1998bu in M~96}

\author[M. Hernandez {\rm et al.}]
{M. Hernandez$^{1}$, 
W.P.S. Meikle$^{1}$,
A. Aparicio$^{2}$, 
C.R. Benn$^{3}$,                
M.R. Burleigh$^{4}$, \and           
A.C. Chrysostomou$^{5}$,                    
A.J.L. Fernandes$^{6}$,            
T.R. Geballe$^{7}$,                         
P.L. Hammersley$^{2}$, \and     
J. Iglesias-Paramo$^{2}$,      
D.J. James$^{8}$,              
P.A. James$^{9}$,               
S.N. Kemp$^{10,2}$,             
T.A. Lister$^{8}$, \and         
D. Martinez-Delgado$^{2}$,      
A. Oscoz$^{2}$,                 
D.L. Pollacco$^{11}$,           
M. Rozas$^{2}$,
S. J. Smartt$^{14}$, \and            
P. Sorensen$^{3}$,             
R.A. Swaters$^{12}$,            
J.H. Telting$^{3}$,             
W.D. Vacca$^{13}$,                
N.A. Walton$^{3}$,\and          
M.R. Zapatero-Osorio$^{2}$      
\\
$^{1}$Astrophysics Group, Blackett Laboratory, Imperial College of
Science, Technology and Medicine, Prince Consort Road, \\ London 
SW7 2BZ \\
$^{2}$ Instituto de Astrofisica de Canarias, 38200 La Laguna,
Tenerife, Spain; and Departamento de Astrofisica, Universidad de \\
La Laguna, Tenerife, Spain \\
$^{3}$Isaac Newton Group, Apartado de Correos 321, 38700 Santa
Cruz de La Palma, Islas Canarias, Spain \\
$^{4}$ Dept. of Physics \& Astronomy, University of Leicester, 
University Road, Leicester, LE1 7RH  \\
$^{5}$Joint Astronomy Centre, 660 N. A'ohoku Place, University Park,
Hilo, Hawaii 96720, USA \\
$^{6}$ Centro de Astrofisica da Universidade do Porto,
Rua das Estrelas, 4150-762 Porto, Portugal. Instituto Superior da Maia, Portugal \\     
$^{7}$ Gemini Observatory, 670 N. A'ohoku Place, Hilo, HI 96720 USA \\
$^{8}$ Dept. of Physics \& Astronomy, University of St. Andrews,
North Haugh, St. Andrews, Fife, KY16 9SS \\
$^{9}$ Astrophysics Research Institute,Liverpool John Moores University,
Twelve Quays House, Egerton Wharf, Birkenhead, L41 1LD \\
$^{10}$ Instituto de Astronomia y Meteorologia, Av. Vallarta No. 2602, Col. Arcos Vallarta, CP 44130, Guadalajara, Jalisco, Mexico \\
$^{11}$ Astrophysics and Planetary Sciences Division, The Queen's University
of Belfast, Belfast BT7 1NN \\
$^{12}$ Kapteyn Institute, Rijksuniversiteit Groningen, Postbus 800, 9700 AV
Groningen, the Netherlands \\
$^{13}$ IRTF, Institute for Astronomy, 2680 Woodlawn Dr., 
Honolulu, HI 96822, USA\\
$^{14}$ Institute of Astronomy, University of Cambridge, Madingley Road, Cambridge, CB3 0HA, UK}

\pagerange{\pageref{firstpage}--\pageref{lastpage}}
\pubyear{1994}

\maketitle

\label{firstpage}

\begin{abstract}
We present first-season infrared (IR) and optical photometry and
spectroscopy of the Type~Ia Supernova~1998bu in M~96.  We also report
optical polarimetry of this event.  SN~1998bu is one of the closest
type~Ia Supernovae of modern times and the distance of its host galaxy
is well-determined.  We find that SN~1998bu is both photometrically
and spectroscopically normal.  However, the extinction to this event
is unusually high, with $A_V=1.0\pm0.11$.  We find that SN~1998bu
peaked at an intrinsic $M_V=-19.37\pm0.23$.  Adopting a distance
modulus of $30.25$ (Tanvir {\it et al.} 1999) and using Phillips {\it
et al.}'s (1999) relations for the Hubble constant we obtain
$H_0=70.4\pm4.3$~km/s/Mpc. Combination of our IR photometry with those
of Jha {\it et al.} (1999) provides one of the most complete
early-phase IR light curves for a SN~Ia published so far.  In
particular, SN~1998bu is the first normal SN~Ia for which good
pre-t$_{Bmax}$ IR coverage has been obtained.  It reveals that the
$J$, $H$ and $K$ light curves peak about 5~days earlier than the flux
in the $B$-band curve.
\end{abstract}


\section{Introduction} 
Supernova~1998bu appeared in the Leo~I group Sab galaxy M~96
(NGC~3368) and was discovered by Mirko Villi (1998) on May~9.9~UT at a
magnitude of about +13 (unfiltered CCD), 10~days before maximum blue
light, t$_{Bmax}$=May~19.8$\pm$0.5~UT (see below).  It is located in
one of the spiral arms and has an offset of 4".3E, 55".3N from M~96's
nucleus (Villi 1998).  The supernova was identified by Ayani {\it et
al.}  (1998) and Meikle {\it et al.}  (1998) as being of type~Ia.  Its
position is RA: 10h 46m 45.95s, Dec:+11$^o$ 50' 07.1" (2000.0).  This
position was determined by us, using a V-band image taken with the
Wide Field Camera of the Isaac Newton Telescope.  Our value agrees
with the positions reported by Nakano \& Aoki (1998) and Boschini
(1998) to within 1.''2.  An $R$-band image of SN~1998bu is shown in
Figure~1.

A pre-discovery observation on May~3.14~UT was reported by Faranda \&
Skiff (1998).  This was about 16.5~days before t$_{Bmax}$ making it
one of the earliest ever observations of a type~Ia supernova (Riess
{\it et al.} 1999).  The Faranda \& Skiff measurement was made with an
unfiltered CCD, and converts to magnitude increments relative to the B
and V maxima of +4.65$\pm$0.18 and +4.75$\pm$0.15 respectively (Riess
{\it et al.} 1999).  These are the largest magnitude increments ever
measured for the rising portion of a type~Ia event.  Riess {\it et
al.} (1999) estimate a B-band rise-time of 19.5$\pm$0.2~days for a
typical SN~Ia.  This implies an explosion date for SN~1998bu of
1998~April~30.2$\pm$0.5~UT.

An important aspect of the discovery of this supernova is that HST
images already exist for its parent galaxy M~96.  Thus, Cepheids
found in these images can be used to find its distance.  Tanvir {\it
et al.} (1995) obtained a distance modulus of $30.32\pm0.15$.
Addition of the 0.05~mag ``long-vs-short exposure'' correction
(Casertano \& Mutchler 1998; Gibson {\it et al.} 2000) increases this
to $30.37\pm0.16$.  This is the value included in Parodi {\it et
al.}'s (2000) estimate of $H_0$ based on SN~Ia observations.  However,
Tanvir {\it et al.}  (1999) have revised their estimate using a larger
number of Cepheids (16 as against 7) in M~96 and applying a correction
for metallicity differences between the LMC and M~96.  They obtain
$30.25\pm0.18$.  Gibson {\it et al.} (2000) use 11 Cepheids in M~96 to
find its distance.  Including their ``typical correction factor'' of
+0.07~mag for metallicity difference between their six principal host
galaxies and the LMC, they obtain a modulus of
$30.27\pm0.10_{rand.}\pm0.16_{sys.}$ for M~96.  This is in good
agreement with Tanvir {\it et al.} (1999).  In this paper, we adopt
the distance modulus of Tanvir {\it et al.} {\it viz.} $30.25\pm0.18$
or a distance of $11.2\pm1.0$~Mpc.  SN~1998bu is one of the closest
type~Ia's of modern times, as well as being observed from an
exceptionally early epoch.

We note that Feldmeier {\it et al.} (1997) estimated the M~96 distance
modulus using the Planetary Nebula Luminosity Function Method (PNLF)
and found a significantly smaller distance modulus of $29.91\pm0.13$.
However, Ferrarese {\it et al.} (2000) have commented on the tendency
of the PNLF method to produce systematically shorter distances than
does the use of Cepheids (or, indeed, the use of the Tip of the Red
Giant Branch method or the Surface Brightness Fluctuation method).  We
therefore do not make use of the Feldmeier {\it et al.} value.

Type~Ia SNe are increasingly recognized as being among the most
reliable indicators of cosmological distances ({\it cf.}  Hamuy {\it
et al.} 1996, Riess {\it et al.} 1996, Perlmutter {\it et al.} 1997).
However, calibration of the zero point requires nearby, well-observed
SNe~Ia at accurately known distances.  Such events are quite rare, but
SN~1998bu is one such example.  A number of major studies of this
event have been carried out.  Suntzeff {\it et al.}  (1999) gave a
detailed description and analysis of the optical light curves acquired
at the Cerro Tololo Inter-American Observatory (CTIO) and Las Campanas
Observatory (LCO). Jha {\it et al.} (1999) presented and discussed
optical and infrared photometry and spectroscopy of the event acquired
at a number of telescopes.

In this paper, we describe IR and optical observations of SN~1998bu
obtained at several telescopes.  A preliminary report of this work is
presented in Meikle \& Hernandez (2000).  In Section~2 we describe our
optical and infrared photometry and spectroscopy, plus optical
polarimetry of SN~1998bu.  In Section~3 the  spectra and light
curves are discussed.  In Section~4 correction for extinction is
determined and the peak absolute magnitudes for SN~1998bu are deduced. We use them to give a value for $H_0$. A brief summary is given in Section~5.

\section{Observations} 
\subsection{Optical Photometry} 
Shortly after the discovery of SN~1998bu we began a programme of $UBVRI$
imaging.  Most of the data were obtained using the 82~cm Instituto de
Astrofisica de Canarias Telescope (IAC80) on Tenerife, and the 1.0~m
Jacobus Kapteyn Telescope (JKT) at La Palma Observatory.  Some
additional photometry was obtained at the 2.56~m Nordic Optical
Telescope (NOT) (La Palma) and the 3.5~m Wyoming-Indiana-Yale-NOAO
telescope (WIYN) at Kitt Peak.  The earliest image was taken on May
13th (JD 2450947.43) at --5~days {\it i.e.}  5~days before the epoch
of maximum blue light, t$_{Bmax}$.  The first season optical
photometry presented here spans 53~days.

The IAC80 observations were acquired with its $1024\times1024$ CCD
camera (scale=0.''433/pixel, fov=$7.3\times7.3$~arcmin.).  Its $BVRI$
central wavelengths are 4500, 5250, 6000 and 8800~\AA.  The JKT
observations were obtained using its $1024\times1024$ CCD camera
(plate scale=0.''331/pixel, fov=$5.6\times5.6$~arcmin.).  Its $UBVRI$
filter transmission characteristics are very close to those of
Johnson-Cousins.  The central wavelengths are, respectively, 3600,
4350, 5350, 6450 and 8400~\AA.  The WIYN photometry was obtained with
a $2048\times2048$ CCD (scale=0.195"/pix, fov=$6.5\times6.5$~arcmin.)
and $UBVRI$ filters centred at 3584, 4327, 5448, 6461 and 8387~\AA\
respectively.  At the NOT, TURPOL was used to obtain photopolarimetry
in $BVRI$, with central wavelengths at 4400, 5300, 5900, and 8300~\AA.

IAC80 data were reduced at the IAC using IRAF software. The
reduction steps included bias-subtraction, flat-fielding and
correction for bad pixels by interpolation. The reduction of the data
from the NOT is described in Oudmaijer {\it et al.} (2000).  Data from
the other telescopes were reduced at Imperial College using the
Starlink package CCDPACK to carry out the standard procedures of
de-biassing, flat-fielding and bad-pixel and cosmic-ray removal.
Aperture photometry was then carried out.  The flux from the supernova
or standard stars was measured in a circular aperture.  The background was
estimated and subtracted using an annulus concentric with the central
aperture.  The annulus inner and outer radii were respectively
$\times$1.5 and $\times$2.5 that of the central aperture.  For a given
night and telescope, the central aperture was selected to have a
diameter equal to four times the FWHM of a typical stellar image.
Owing to its spatial variation, particular care had to be taken in
estimating and subtracting the background from the host galaxy.  To
check this, we examined $VRI$ images of M~96 taken by N. Tanvir with
the INT in 1994.  We found that variation of the aperture annulus
radius from $\sim$10'' to $\sim$20'', centred on the supernova
position, would affect the supernova instrumental magnitudes presented
here by no more than 0.01 mag.  Instrumental magnitudes were then
obtained using the Starlink PHOTOM package. Photometry was performed
in two steps.  First, the supernova magnitudes relative to comparison
field stars were measured.  The comparison stars were then calibrated
against Landolt field stars.  The comparison stars are identified as
CS1, CS2, CS3, CS4 and CS5 in Figure~1. At least three of these were
usually available on a given frame.

For the IAC80 observations, colour-corrected magnitude differences
between the comparison stars and SN~1998bu were obtained using the
canonical equations (1) {\it viz.}:
\begin{eqnarray*}
\\
\Delta b = \Delta b_{o} + C_{b} \Delta(b_{o}-v_{o})
\\
\Delta v = \Delta v_{o} + C_{v} \Delta(b_{o}-v_{o}) & (1)
\\
\Delta r = \Delta r_{o} + C_{r} \Delta(v_{o}-r_{o})
\\
\Delta i = \Delta i_{o} + C_{i} \Delta(r_{o}-i_{o})
\end{eqnarray*}
where $\Delta b_{o}$, $\Delta v_{o}$, $\Delta r_{o}$, $\Delta i_{o}$
are the instrumental magnitude differences between a comparison star
and the supernova, $\Delta(b_{o}-v_{o})$, $\Delta(v_{o}-r_{o})$,
$\Delta(r_{o}-i_{o})$ are the colour differences, and $C_{b}$,
$C_{v}$, $C_{r}$, $C_{i}$ are the colour coefficients, derived below
using the procedures of Hardie (1962).  No airmass term appears since,
for any frame, the comparison stars and supernova were observed at
essentially the same airmass.  The comparison stars were calibrated in
$BVRI$ using standard Landolt (1992) fields.  The observations for
this were carried out with the IAC80 on 1999 June 13.  The colour
correction coefficients were also derived from these data using 
IRAF's PHOTCAL package. The respective values for ($C_{b}$, $C_{v}$,
$C_{r}$, $C_{i}$) were (0.022$\pm$0.010, 0.014$\pm$0.007,
0.036$\pm$0.015, 0.075$\pm$0.020).  These values agree well with those
shown in the IAC80 web-page
(http://www.iac.es/telescopes/iac80/instrumentacion.html\#color).

For the WIYN observations, calibration of the comparison stars in
$VRI$ was carried out using Landolt standards observed at the WIYN
telescope on 1998 June 5.  Similar procedures were used as for the
IAC80 calibrations.  However, no Landolt standards were observed in
$U$ or $B$.  Therefore we used a modified version of equations (1)
{\it viz.}:
\begin{eqnarray*}
\\	 
    \Delta v = \Delta v_{o} + C'_{v} \Delta (v_{o}-r_{o})
\\
    \Delta r = \Delta r_{o} + C'_{r} \Delta (v_{o}-r_{o}) & (2)
\\
    \Delta i = \Delta i_{o} + C'_{i} \Delta (r_{o}-i_{o})
\end{eqnarray*}
Values for ($C'_{v}$, $C'_{r}$, $C'_{i}$) were found to be
(0.002$\pm$0.008, -0.014$\pm$0.018, 0.073$\pm$0.028).

The $VRI$ comparison star magnitudes obtained from the IAC80 and WIYN
all agreed to within the errors, and so the weighted mean values
from these telescopes were adopted. The $BVRI$ comparison
star magnitudes are shown in Table~1.

Unfortunately, in the case of the JKT the observations that were
obtained of standard star fields did not span an adequate colour
range.  Moreover, the constraints of scheduled JKT observers meant
that often only two or three filters were available, and in a variety
of combinations.  In view of this, we did not carry out colour
corrections for the JKT data.  Colour correction procedures, such as
those described above, can alter the magnitude by as much as 0.1, with
the $U$ and $B$ filters usually being the most sensitive to this
effect.  However, since our SN~1998bu magnitudes were obtained by
averaging the values obtained relative to comparison stars of
different colour indices, we expect the error due to inadequate colour
correction to be small.  Nevertheless, for such cases, we have
increased the uncertainty to $\pm$0.075.

We then used the calibrated magnitudes of the comparison stars
(Table~1) to transform the colour-corrected differences from equations
(1) or (2) into apparent magnitudes for SN1998bu using equations (3) {\it
viz.}:
\begin{eqnarray*}
\\
    B_{sn} = B_{cs}-\Delta b
\\
    V_{sn} = V_{cs}-\Delta v & (3)
\\
    R_{sn} = R_{cs}- \Delta r 
\\
    I_{sn} = I_{cs}- \Delta i
\end{eqnarray*}

Only a few $U$-band measurements were obtained.  Moreover, due to the
lack of observations of standard stars in this band, we had to
indirectly calibrate the comparison stars.  To achieve this, we used
average $U$-band magnitudes for these stars from Jha {\it et al.}
(1999) and Suntzeff {\it et al.} (1999).  Given the larger
uncertainties in this procedure we estimate a precision of no better
than $\pm$ 0.1 in the $U$-band apparent magnitudes of ~SN1998bu.

Thus, at a given epoch and telescope, for each available comparison
star a set of magnitudes was calculated for SN~1998bu.  The weighted
mean magnitude in each band was then obtained. These are listed in
Table~2.

\begin{table*}
\centering
\caption[]{Magnitudes of Comparison Stars near SN1998bu}
 \begin{minipage}{\linewidth}
\begin{tabular}{rcccc} \hline
Star    & B       &V       &R       &I       \\ \hline
CS1  &13.607(18)\footnote{Figures in brackets give the internal error of each magnitude and are quoted in units of the magnitude's least significant digit}  &13.092(10) &12.776(10) &12.481(13)  \\
CS2  &15.559(40) &15.039(15) &14.712(22) &14.393(25) \\
CS3  &17.320(93) &16.509(30) &15.867(48) &15.277(55) \\
CS4  &16.410(42) &15.807(16) &15.369(25) &14.948(30) \\
CS5  &15.497(36) &14.898(12) &14.531(13) &14.180(18)  \\
\hline
\end{tabular} \\
\end{minipage}
\end{table*}

\begin{table*}
\centering
\caption[]{Optical photometry of SN1998bu}
\begin{minipage}{\linewidth}
\renewcommand{\thefootnote}{\thempfootnote}
\begin{tabular}{cccccccll} \hline
Julian Day & Epoch\footnote{Relative to t$_{Bmax}$ = 1998 May 19.8 UT.} & U    & B    & V  & R    & I    & Telescope & Observer    \\ 
 (2450000+)  & (d) &&&&&&&\\ \hline
947.43& --5.87&           &12.502(28)\footnote{Figures in brackets give the internal error of each magnitude and are quoted in units of the magnitude's least significant two digits.}&    12.199(14)&    11.848(16)&     11.727(20)&  IAC80 & A. Oscoz  \\ 
948.38& --4.92&           & 12.401(75)&     &            11.808(75)&     &          JKT&                  P. James  \\
948.43& --4.87&           & 12.403(28)&    12.117(14)&    11.818(15)&     11.676(19)&  IAC80&             A. Oscoz  \\
949.37& --3.93&           & 12.340(75)&     &            11.760(75)&     &          JKT&                  P. James  \\
949.40& --3.90& 11.99(20)& 12.300(40)&    12.070(30)&    11.770(40)&    11.660(50)&  NOT&                EXPORT\footnote{EXtra-solar Planet Observational Research Team} \\
950.41& --2.89& 11.99(20)& 12.250(50)&    &            11.770(50)&    11.670(50)&  NOT&                  EXPORT \\
950.42& --2.83&           & 12.276(75)&     &            11.727(75)&     &          JKT&                  P. James  \\
951.37& --1.93&           & 12.260(75)&    &            11.698(75)&     &          JKT&                   P. James  \\
952.36& --0.94&           & 12.239(75)&     &            11.704(75)&     &          JKT&                  P. James  \\   
953.37& +0.07&           & 12.239(75)&     &            11.684(75)&     &          JKT&                   P. James  \\ 
954.37& +1.07&           & 12.250(75)&     &            11.682(75)&     &          JKT&                   P. James  \\
955.36&  +2.06&          & 12.265(75)&  11.881(75)& 11.713(75)& 11.827(75)&       JKT&                    M. Burleigh \& S. Smartt \\
956.40&  +3.10&          & 12.326(75)&  11.907(75)& 11.721(75)& 11.859(75)&       JKT&                    M. Burleigh \& S. Smartt \\
957.39& +4.09&12.20(10) & 12.370(75)&   11.960(75)& 11.739(75)& 11.905(75)&       JKT&                    M. Burleigh \& S. Smartt \\
958.36& +5.06&12.30(10) & 12.329(75)&  11.986(75)&11.809(75)&11.988(75)&          JKT&                    M. Burleigh \& S. Smartt \\
958.38& +5.08&           & 12.423(16)&   11.962(19)& 11.777(20)& 11.912(33)&     IAC80                  &  E. Gonzalez  \\
959.40& +6.10& 12.41(10) & 12.470(75)&   11.989(75)& 11.858(75)& 12.055(75)&       JKT&                   M. Burleigh \& S. Smartt \\
963.66&  +10.36&          &12.817(26)& 12.298(12)& 12.249(21)& 12.287(30)&      WIYN&                     D. Harmer  \\
964.39&  +11.09&          &12.880(75)&  12.298(75)& 12.261(75)& 12.327(75)&      JKT&                       R. Swaters \\
965.66&  +12.36&          &12.920(35)& 12.419(16)& 12.350(24)&12.352(25)&       WIYN&                     D. Harmer  \\
966.38&  +13.08&          &13.104(75)&  12.448(75)& 12.398(75)& 12.398(75)&       JKT&                    R. Swaters \\
968.38&  +15.08&          &13.352(32)& 12.558(19)& 12.414(25)& 12.332(26)&      IAC80&                    S. Kemp \& P. Hammersley    \\
968.72&  +15.42&13.44(10)& 13.367(27)& 12.609(12)& 12.481(16)& 12.309(18)&      WIYN&                     D. Harmer  \\
969.34&  +16.04&          &13.381(75)& 12.644(75)& 12.419(75)&   &          JKT&                          D. James   \\
969.42&  +16.12&          & 13.498(36)& 12.616(20)&  12.431(24)& 12.284(25)&     IAC80&                   S. Kemp    \\
969.70&  +16.40&13.76(10)& 13.480(31)& 12.650(14) & 12.480(15) & 12.275(20)&    WIYN&  D. Harmer \\
970.34&  +17.04&          &13.487(75)& 12.719(75)& 12.443(75)&    &             JKT&                      D. James   \\
971.33&  +18.03&          &13.607(75)& 12.784(75)& 12.444(75)&   &                JKT&                    D. James   \\ 
972.35&  +19.05&          &13.717(75)& 12.834(75)& 12.452(75)&   &              JKT&                      D. James   \\
972.68&  +19.38&14.16(10) &13.821(28)& 12.791(12)&  12.489(17)&  12.200(18)&        WIYN&                 D. Harmer  \\
973.33&  +20.03&            &13.782(75)& 12.888(75)& 12.465(75)&   &           JKT&                       D. James   \\
974.33&  +21.03&            &13.912(75)& 12.958(75)& 12.472(75)&   &            JKT&                      D. James   \\
976.39&  +23.09&14.60(10) &14.218(75)   & 12.955(75)&  12.494(75)&  12.096(75)&     JKT&                  J. Telting \\
977.45&  +24.15&14.72(10) & 14.316(76)& 13.035(19)&  12.532(24)&  12.102(25)&     IAC80; U-JKT&                  M. Rozas   \\
978.39&  +25.09&            &14.461(74)& 13.055(18)&  12.563(24)&  12.077(27)&     IAC80&                 M. Rozas   \\    
981.34&  +28.04&15.20(10) & 14.649(75)& 13.246(75)&   &          12.101(75)&     JKT&                     A. Fernandes  \\
983.39&  +30.09&15.22(10)  &14.867(75)& 13.461(75)&  12.804(75)&  12.184(75)&     JKT&                    A. Fernandes \\        
985.35&  +32.05&            &14.885(75)&      &        &          12.473(75)&     JKT&                    D. Martinez \& A. Aparicio \\  
985.40&  +32.10&            &15.039(50)& 13.530(18)&  12.899(25)&  12.403(25)&     IAC80&                 E. Gonzalez \& M. Zapatero-Osorio \\
986.34&  +33.04&            &14.928(75)&     &        &          12.473(75)&     JKT&                     D. Martinez \& A. Aparicio\\
990.34&  +37.04&            &15.135(75)&     &        &            &           JKT&                       D. Martinez \& A. Aparicio\\
997.40&  +44.10&            &15.286(68)& 13.990(19)&  13.578(20)&  13.141(26)&     IAC80&                 A. Oscoz    \\
1000.40& +47.10&            &15.372(70)& 14.097(21)&  13.677(25)&  13.292(28)&     IAC80&                 S. Kemp     \\
\hline
\end{tabular} \\
\end{minipage}
\end{table*}

\subsection{Infrared Photometry} 
SN~1998bu yielded one of the earliest sets of {\it near-IR} photometry
ever obtained for a supernova.  Indeed, this is the first time that IR
photometry for a normal type~Ia event has been acquired {\it before}
t$_{Bmax}$.  The earliest IR observation was by Mayya \& Puerari
(1998) who acquired JHK photometry at --8.4~days using the 
Observatorio Astronomico Nacional 2.1-m telescope (+CAMILA/NICMOS 3
camera) at San Pedro Martir, Mexico.  During the first season of SN~1998bu
observations, IR photometry was acquired at a number of telescopes.
Some of these data have been published in Jha {\it et al.} (1999).
Preliminary IR light curves were also displayed in Meikle \& Hernandez
(2000).  Here, we present a  description of IR photometry
carried out at 1.5m  Telescopio ``Carlos Sanchez'' (TCS), Tenerife, the
University of Hawaii 2.2m Telescope (UH2.2), Hawaii, the 3.8m United
Kingdom Infrared Telescope (UKIRT), Hawaii and the 4.2m William Herschel
Telescope (WHT), La Palma.  The data are presented in Table~3.

\begin{table*}
\centering
\caption[]{Infrared photometry of SN1998bu}
\begin{minipage}{\linewidth}
\renewcommand{\thefootnote}{\thempfootnote}
\begin{tabular}{cccccll} \hline
Julian Day & Epoch\footnote {Relative to t$_{Bmax}$ = 1998 May 19.8 UT.} & $J$  &    $H$    &    $K$    & Telescope & Observer \\
(2450000+)  & (d) &&&&& \\ \hline
949.40& --3.8 & 11.55(3)\footnote{Figures in brackets give 
the internal error of each magnitude and are
quoted in units of the magnitude's least significant digit}
 & 11.59(3) & 11.42(3) &   TCS     & EXPORT team   \\
950.40& --2.8 & 11.68(3) & 11.86(3) & 11.44(3) &   TCS     & EXPORT team   \\
952.74& --0.5 & 11.71(4) & 11.88(4) & 11.63(3) &   UKIRT   & A. Chrysostomou \\
953.40&   0.2 & 11.89(5) & 11.95(5) & 11.60(5) &   TCS     & P. Hammersley \\
955.40&   2.2 & 11.87(5) & 11.83(5) & 11.66(5) &   TCS     & P. Hammersley \\
957.40&   4.2 & 12.06(5) & 11.88(5) & 11.61(5) &   TCS     & P. Hammersley \\
958.40&   5.2 & 12.05(5) & 11.96(5) & 11.75(5) &   TCS     & P. Hammersley \\
959.84&   6.6 & 12.42(3) & 11.97(3) & 11.88(3) &   UKIRT   & T. Geballe    \\
970.35&  17.2 &    ---   & 11.75(5) & 11.84(4) &   WHT     & C. Benn       \\
976.77&  23.6 & 13.08(1) & 11.74(2) & 11.92(2) &   UH2.2   & W. Vacca      \\ \hline 
\end{tabular} \\
\end{minipage}
\end{table*}

All the TCS data were taken with the CVF IR photometer.
The photometric system has been accurately characterized (Alonso {\it et al.}, 1994) and is very similar to the UKIRT system. For the four epochs in which the observations were carried out, a 20~arcsec. aperture
and a chop throw of about 35 arcsec. at 6.7 Hz were employed.
Conditions were judged to be photometric on all four nights, and that
the photometry accuracy was no worse than $\pm$3\%.  The data were
reduced using the IAC data reduction program TCSPHOT. Absolute
calibration was by means of repeat measurements of a sequence of
standard stars at a range of air-masses.  The stars were BS3304,
BS4039, BS4883, BS5384, BS5423A and BS6538A.  Error estimates are
based on the statistical error in the measurement of the supernova
together with a smaller contribution from the measurement of the
calibration sources.  TCS observations were also made on 2 successive
nights by the EXtra-solar Planet Observational Research Team (EXPORT)
during the 1998 international time of the Canary Islands
Observatories. 
        
The UH2.2 observations were made with the QUIRC camera,
which contains a 1024x1024 HgCdTe array. The plate scale is 0.189"/pixel.
A 7~point dither pattern was used, chopping to sky after each
individual exposure. Sky frames and flat fields were made from these
offset sky fields. The airmass ranged between 1.2 and 1.3. The
night was photometric, and observations of the following standard
stars were made to compute the magnitudes: FS21, FS23, FS27, and FS35.
The magnitudes of the FS stars (as well as the additional stars in the
fields of FS23 and FS35) given by Hunt {\it et al.} (1998) were used for
transforming between instrumental and intrinsic magnitudes. 
Instrumental magnitudes were computed using an aperture of
15 pixels in radius, with a background aperture of 20--30 pixels.  The
seeing was typically 0.7--0.8".

The UKIRT observations were made with the IRCAM3 camera which uses a
256x256 InSb array.  The plate scale was 0.286''/pixel and the seeing
was typically 1.5''.  The WHT observations were made with the WHIRCAM
camera which also uses a 256x256 InSb array.  The plate scale is
0.240''/pixel and the seeing was typically 0.8-0.9''. For both the
UKIRT and WHT observations a 5 point dither pattern was used.  The
data were reduced using the package IRCAMDR (Aspin 1996).  Calibration
of the UKIRT data was by means of the standard stars HD84800 (19 May)
and FS20 (26 May), and the WHT data by means of FS25.  For the UKIRT
and WHT images, instrumental magnitudes were computed by increasing
the aperture size until the supernova to standard ratio converged to a
constant value ($\pm$1\%).  This was usually attained for aperture
radii of 10--15 pixels.

\begin{table*}
\centering
\caption[]{Polarimetry of  SN1998bu (EXPORT team)}
\begin{minipage}{\linewidth}
\renewcommand{\thefootnote}{\thempfootnote}
\begin{tabular}{ccccccc} \hline
Julian Day & Epoch\footnote{Relative to t$_{Bmax}$ = 1998 May 19.8 UT.} & Band & POL  &   Error    & Posn. Angle      & Error \\
(2450000+) & (d) &&(\%)&(\%)&(deg.)& (deg.)\\
\hline
949.40& --3.90 & U & 1.227 & 0.076 & 186.3 & 1.8 \\
949.40& --3.90 & B & 1.597 & 0.072 & 184.6 & 1.3 \\
949.40& --3.90 & V & 1.695 & 0.105 & 179.1 & 1.8 \\
949.40& --3.90 & R & 2.101 & 0.068 & 180.3 & 0.9 \\
949.40& --3.90 & I & 1.871 & 0.145 & 173.4 & 2.2 \\
\\
950.41& --2.89 & U & 1.091 & 0.075 & 184.2 & 2.0 \\
950.41& --2.89 & B & 1.469 & 0.074 & 181.2 & 1.4 \\
950.41& --2.89 & V & 1.842 & 0.101 & 177.8 & 1.6 \\
950.41& --2.89 & R & 1.894 & 0.065 & 177.9 & 1.0 \\
950.41& --2.89 & I & 1.788 & 0.107 & 174.5 & 1.7 \\
\\
951.40& --1.90 & U & 1.222 & 0.079 & 182.3 &1.9  \\
951.40& --1.90 & B & 1.689 & 0.073 & 181.0 & 1.2 \\
951.40& --1.90 & V & 1.757 & 0.108 & 178.6 & 1.8 \\
951.40& --1.90 & R & 1.965 & 0.105 & 179.5 & 1.5 \\
951.40& --1.90 & I & 1.668 & 0.129 & 176.5 & 2.2 \\ \hline
\end{tabular}
\end{minipage}
\end{table*}

\subsection{Polarimetry}
During the nights of 15-17 May 1998, $UBVRI$ polarimetric observations
were taken of SN~1998bu using the Turpol instrument, mounted on the
2.5~m Nordic Optical Telescope (NOT), La Palma, Spain.  The
observations were carried out by EXPORT members and data-reduction are discussed in detail in Oudmaijer
{\it et al.} (2000).

The light from the direction of SN~1998bu was relatively strongly
polarized (see Table~4).  We wish to decide whether the polarization
is intrinsic to the source, due to Galactic dust extinction, or
extinction within M~96 itself.  Time-variability in the polarization
would have proven that it was intrinsic to the source, but no
variation was detected over the three nights.  Field stars near to
M~96 showed little degree of polarization.  This is consistent with
the dust maps of Schlegel {\it et al.} (1998) which indicate that the
extinction contribution from our Galaxy is small.  It appears,
therefore, that most of the polarization must have been produced
within M~96.  This is important, since the polarization as a function
of wavelength behaved similarly to that seen for normal interstellar
polarization {\it i.e.}  the data follow the Serkoswki model
(R. Oudmaijer, personal communication). This indicates that the ISM of
M~96 is somewhat similar to that of the Milky Way.  In Section~3, we
show that SN~1998bu is highly reddened and that most of this reddening
probably arises from dust extinction within M~96.  It therefore seems
likely that observed polarization is associated with interstellar
material within M~96.

\subsection{Optical Spectroscopy} 
Optical spectroscopy was acquired using the ISIS spectrograph of the
William Herschel Telescope (WHT), the IDS spectrograph of the Isaac
Newton Telescope (INT) and the Hydra spectrograph of the
Wisconsin-Indiana-Yale-NOAO Telescope (WIYN). The log of observations
is given in Table~5.  The spectra were reduced by means of the package 
~FIGARO (Shortridge {\it et al.}, 1995). Debiasing and flat-fielding were
carried out in the usual manner.  Wavelength calibration was by means
of arc lamp spectra, and the uncertainty was typically less than
$\pm$1~\AA.  The spectra were relatively fluxed by comparison with the
flux standard Feige~34. However, absolute fluxing was more difficult
owing to variable observing conditions which resulted in uncertain
amounts of vignetting of the target and standard by the slit.  To
correct for this systematic error, we used the $BVR$ magnitudes
obtained from the JKT images.  Transmission functions for the $B$, $V$
and $R$ bands were constructed by multiplying the JKT filter functions
by the CCD response and the standard La Palma atmospheric transmission
function.  The relatively-fluxed spectra were then multiplied by the
net transmission functions, and the resulting total flux within each
band compared with JKT-derived magnitudes corresponding to the same
epochs.  Thus, correction (scaling) factors were obtained for each
spectrum.  Apart from the earliest spectrum (--6.8~d), $BVR$
magnitudes corresponding to the spectroscopy epochs were obtained
either from actual simultaneous observations or by interpolation
within the JKT data.  Fluxing of the --6.8~d spectrum was less certain
as it was obtained 2~days before the earliest JKT photometry point.
We therefore used photometry gleaned from the IAU circulars and from
Suntzeff {\it et al.} (1999) and Jha {\it et al.} (1999) to estimate
the $BVR$ magnitudes at --6.8~d.

Scaling factors ranged from $\times$0.83 to $\times$2.23.  The scaling
factor closest to unity ($\times$0.94) was for the --3.8~d spectrum.
This was expected since the flux scale for this spectrum had already
been corrected using a low-resolution spectrum taken through a very
wide (7~arcsec.) slit.  For a given epoch, the scaling factors for
each band agreed to within $\pm$5\% demonstrating good internal
consistency for the procedure.  The relatively-fluxed spectra were
therefore multiplied by the geometric mean of the scaling factors for
each epoch.  Including the uncertainty in the photometry, we estimate
that the fluxing accuracy is better than $\pm$10\% except for --6.8~d
where the error is probably closer to $\pm$15\%.  The optical spectra
are shown in Figure~2.  The --6.8~d spectrum is the earliest reported
for SN~1998bu.

\begin{table*}
\centering
\caption{Log of optical and infrared spectroscopy of SN1998bu}
\begin{minipage}{\linewidth}
\renewcommand{\thefootnote}{\thempfootnote}
\begin{tabular}{ccccccl} \hline
Julian Day 
& Date UT\footnote{{\it Start} time for integrations on SN 1998bu.} 
& Epoch\footnote{Relative to t$_{Bmax}$ = 1998 May 19.8 UT.} 
& Telescope/Instrument & $\lambda\lambda$ 
& $\Delta\lambda$\footnote{Spectral resolution. The $\Delta\lambda$ 
values for UKIRT/CGS4 are for 10,000~\AA\ and 16,500~\AA\ respectively.} 
& Observer \\ 
(+2450000) & (1998) &(d)&& (\AA) &(\AA) &\\ \hline
946.40 & May 12.90 & --6.8 & WHT/ISIS   & 3590--9097 & 5.8 & J. Iglesias-Paramo \\
949.38 & May 15.88 & --3.8 & INT/IDS    & 3300--9072 & 13  & P. Sorensen\footnote{Observations carried out during EXPORT telescope time} \\
951.40 & May 17.90 & --1.8 & INT/IDS    & 3300--9081 & 13  & EXPORT \\
952.40 & May 18.90 & --0.8 & INT/IDS    & 3207--9460 & 6.6 & D. Pollacco \\
964.72 & May 31.22 & +11.5 & WIYN/Hydra & 3541--5573 & 4.0 & P.S. Smith     \\
966.68 & June 2.18 & +13.5 & WIYN/Hydra & 4910--10265& 5.2 & D. Willmarth \\ 
972.40 & June 7.90 & +19.2 & INT/IDS    & 3602--9298 & 6.6 & D. Pollacco \\ 
988.75 & June 24.25 & +35.5 & UKIRT/CGS4 & 8175-20966 & 12.5 \& 41 & T. Geballe \\ \hline
\end{tabular} 
\end{minipage}
\end{table*}

\subsection{Infrared Spectroscopy} 
IR spectroscopy at UKIRT was obtained on 1998 June~24.2~UT (+35.5~d)
using CGS4, its 40 l/mm grating and a 0.61" (1~pixel) wide slit (see
Table~5).  The spectra were sampled every 1/2 pixel; the resolution 
of the spectrometer was approximately 370~km/s in the $IJ$-bands
and 450~km/s in the $H$~band.  During the observations the telescope
was nodded 7.5~arcsec. along the slit.

The spectra were reduced using the standard procedures of the package 
FIGARO, including optimal extraction (Horne, 1986). Wavelength
calibration was by means of an arc spectrum, and is judged to be
accurate to better than 2~\AA\ in the IJ-bands and 3~\AA\ in the
H-band.  Relative fluxing was by comparison with spectra of BS~4281
($IJ$-bands) and BS~4079 ($H$-band).  For BS~4281 (F5V) we assumed
$J$=+5.845 and a temperature of 6540~K.  For BS~4079 (F6V) we assumed
$J$=+5.770 and a temperature of 6450~K.  Final absolute fluxing was
achieved using the composite $J$- and $H$-band light curves described
in section 3.2.  From these we obtain $J=+13.18\pm0.10$ and
$H=+12.24\pm0.10$ at the epoch of the IR spectra.  Magnitudes were
then derived from the IR spectra using the combined filter passband
plus atmospheric transmission response functions provided on the UKIRT
web pages together with the absolute spectrum of Vega. On comparison
with the light curve-derived values it was found that the
spectrum-derived magnitudes were too faint by factors of $\times$1.02
in $J$ and $\times$1.23 in $H$. The fluxes of the spectra were
therefore multiplied by these factors.  The $I$-band spectrum
overlapped the $J$-band spectrum in the 10,000--11,000~\AA\ region. We
multiplied the $I$-band flux by $\times$1.14 to bring it into
agreement with the $J$-band.  We believe the final fluxing is accurate
to $\pm$15\%. The IR spectrum is displayed in Figure~3. It spans
8,175--20,966~\AA.  (The short wavelength coverage of CGS4 now
overlaps the typical long wavelength limit of optical spectrographs,
providing access to the poorly explored 0.9--1.0~$\mu$m region.)

\section{Results}
\subsection{Spectra}
\subsubsection{Optical Spectra}
The optical spectra demonstrate that SN~1998bu was a spectroscopically
normal, but highly reddened type~Ia supernova. This is illustrated in
Figure~4 where we compare the spectra of SN~1998bu at maximum light
with those of the normal type~Ia SNe~1981B (Branch {\it et al.}, 1983)
and 1994D (Meikle {\it et al.}, 1996).  The three spectra are quite
similar. The main difference is due to the greater reddening of
SN~1998bu (see below).  Redward of 5000~\AA, SN~1998bu and SN~1981B
have greater similarity, although the calcium triplet absorption
around 8250~\AA\ is considerably deeper in SN~1981B.  At shorter
wavelengths the fine structure of the SN~1998bu spectrum are generally
closer in appearance to those of SN~1994D.  Jha {\it et al.} (1999)
pointed out the existence of an unidentified absorption feature
blueward of the CaII H \& K absorption, and suggested it could be due
to silicon or calcium. This feature is clearly visible at $\sim$3,700
\AA\ in all the SN~1998bu spectra up to $t_{Bmax}$, including the
first spectrum at -6.8d (Figure~2).  The feature is not present in the
maximum light spectrum of SN~1981B but is very strong in SN~1994D at
the same epoch (Figure~4).  However, later spectra (day~+9 (Jha {\it
et al.}; day~+11.5 (this work)) show the feature has weakened
considerably.  By day~+19.2d it has essentially vanished. \\

\subsubsection{Infrared Spectrum}
The strong P~Cygni feature at 8,175--8,700~\AA\ is due to the calcium
triplet (Filippenko 1997). At $\sim$10,000~\AA\ there is a
particularly prominent, isolated feature.  In the rest frame of the
host galaxy the feature peaks at 9950$\pm$150~\AA\ and has a FWHM
equivalent to $\sim$8,000~km/s.  If we interpret the trough to the
blueward side as being the absorption component of a P-Cygni profile
then the blueshift of the trough is 8,170$\pm$570~km/s.
P. H{\"o}flich (private communication) suggests that this line can be
identified with the very strong Fe~II z$^4$F$_4$--b$^4$G$_5$
9,997.56~\AA\ line.  This line is predicted in some of the model
spectra of Wheeler {\it et al.} (1998). Between 10,000 and 12,000~\AA,
SN~1998bu exhibits the dramatic drop responsible for the typical red
J--H colour of type~Ia events at this time.  Many of the other
features in the IR spectrum are probably due to singly and
doubly-ionized cobalt and iron (Bowers {\it et al.}  1997).

In Figures~5 \& 6 we compare the IR spectrum of SN~1998bu with those
of other SN~Ias over a range of epochs (Bowers {\it et al.} 1997). The
spectrum nearest in epoch to the SN~1998bu +35.5~d spectrum is the
+40~d spectrum of SN~1992G. The two spectra are similar.  It can be
seen that the prominent Fe~II feature at $\sim$10,000~\AA\ is quite
common in early time type~Ia events, and persists from as early as 
+20~days to as late as +60~days.  However, by +92 days the feature
appears to weaken.  It could prove to be a valuable line for
determination of both abundance and velocity distribution
(P. H{\"o}flich, private communication.)

\subsection{Light Curves} 
\subsubsection{Optical Light Curves} 
The optical photometry (Table~2) is plotted as light curves in
Figure~7.  The shapes of the light curves are typical for a normal
type~Ia supernova, and agree with those presented by Suntzeff {\it et
al.} (1999) and Jha {\it et al.} (1999).  The $BRI$ magnitudes and
epochs at maximum light were estimated by fitting low order
polynomials. In the $V$~band, the light curve is not well sampled
around maximum.  We therefore fitted the $V$ template of Leibundgut
(1988) to find the epoch of maximum light in this band.  The values
are shown in Table~6.  Maximum light in the $B$-band occurred on
May~19.8$\pm$0.5 days (UT) and we adopt this as the fiducial
t$_{Bmax}$=0~days.  Suntzeff {\it et al.}  (1999) found t$_{Bmax}$ to
be on May~19.4$\pm$0.5 (UT) and Jha {\it et al.}  obtained
May~19.3$\pm$0.8~days. Thus all three estimates agree to within the
uncertainties.

From our $B$-band light curve we find the decline rate parameter
$\Delta$$m_{15}$(B)=1.06$\pm$0.05.  This is consistent with the
observed values yielded by the $B$-band light curves of Suntzeff {\it
et al.} (1999) and Jha {\it et al.} (1999).  Phillips {\it et al.}
(1999) point out that the decline rate is a weak function of
extinction. In their Equ.~6 they provide a correction relation which
we find yields a correction of +0.03 for SN~1998bu (see also section 3.4). We therefore
adopt $\Delta$$m_{15}$(B)=1.09$\pm$0.05.

\begin{table*}
\centering
\caption[]{Epochs and magnitudes at maximum light}
\begin{minipage}{\linewidth}
\renewcommand{\thefootnote}{\thempfootnote}
\begin{tabular}{cccccc} \hline
Julian Day & Epoch\footnote{Relative to t$_{Bmax}$ = 1998 May 19.8 UT.} & Waveband & Apparent Mag. & De-reddened Mag. & Absolute Mag. \\
(+2450000) & (d) &&&&\\ \hline 
953.3(0.5) & 0.0 & B & 12.24(08)\footnote{Figures in brackets give the
internal errors in units of the least significant two digits.} & 10.90(17)\footnote{The error given in this column is the combination of the random photometric error and the systematic error in  the reddening correction.}& --19.35(25)\footnote{The error given in this column is the combination of the random photometric error and the systematic error in  the distance and reddening correction.}  \\ 
955.0(1.0)& +1.7 & V & 11.88(10) & 10.88(14) & --19.37(23) \\ 
953.6(0.5)& +0.3 & R & 11.68(08) & 10.93(11) & --19.32(21) \\ 
950.3(0.5) & --3 & I & 11.66(08) & 11.18(10) & --19.07(21) \\ 
948.0(2.0)&  --5 & J & 11.55(25) & 11.27(25) & --18.98(31) \\ 
948.0(2.0)&  --5 & H & 11.60(25) & 11.41(25) & --18.84(31) \\ 
948.0(2.0)&  --5 & K & 11.40(25) & 11.29(25) & --18.96(31) \\ 
\hline
\end{tabular} \\
\end{minipage}
\end{table*}

The $VRI$ fluxes peaked at, respectively, $+1.7\pm1.1$~d,
$+0.3\pm0.7$~d,  and $-3.0\pm0.7$~d.  This is consistent with Suntzeff
{\it et al.} (1999) who report maxima at +1.2$\pm$0.7, +0.6$\pm$0.7
and --3.4~$\pm$1.1~days for $VRI$ respectively, and with Jha {\it et
al.} (1999) who find the V-maximum to have occurred at
+1.6$\pm$1.3~days.  This behavior has been noted in other type~Ias
such as SN~1990N and SN1992A (Suntzeff 1993, Leibundgut 1998, Lira
{\it et al.} 1998).  Clearly, this is not the behavior of a simple
cooling blackbody, where the maximum would occur later at longer
wavelengths. That the photosphere is not a pure blackbody is
confirmed by the contemporary spectra (Figure~2).  We note that for
SN~1998bu the times between maxima in different bands do {\it not}
agree with the analysis of SN~Ia light curves by Schlegel (1995).  He
obtained $t_{Rmax}$-$t_{Imax}~+1.6$~d, $t_{Rmax}$-$t_{Vmax}~+3.6$~d
and $t_{Imax}$-$t_{Vmax}~+2.0$~d, while our results for SN~1998bu are
+3.3, --1.4 and --4.7~d respectively.  We also see a pronounced second
maximum in $I$ at about +25~d together with a corresponding inflection
in $R$.  Again, this behavior has been seen in other type~Ias ({\it
e.g.} Ford {\it et al.} 1993).

\subsection{Infrared Light Curves.}
In Figure~8 we show both the optical and infrared light curves.  The
latter were obtained by plotting our IR photometry (Table~3) together
with data from Mayya \& Puerari  (1998) and Jha {\it et al.} (1999).
These data constitute one of the most complete early-time infrared
light curves obtained for a type~Ia supernova.  As mentioned
above, it is the first time that IR photometry for a normal type~Ia
event has been acquired {\it before} t$_{Bmax}$. We find that the {\it
first} maximum in the IR light-curves occurs at about -5~d.  Thus,
there is a trend in which the epochs of first maximum occur earlier as
we move from the $R$-band through $I$ and into the IR.  We also show
in Figure~8 the $JHK$ template light curves of Elias {\it et al.}
(1985). (We have slightly truncated Elias {\it et al.}'s original
templates so that the earliest epoch of the template corresponds to
Elias {\it et al.}'s earliest observation.)  The position of these
templates were fixed on the time axis assuming that the fiducial time,
$t_0=0$, of Elias {\it et al.}  corresponds to ~-6.25~days.  The IR
light curves have been shifted vertically to provide the best match to
the data.  A detailed discussion of the IR light curves is given in
Meikle (2000).

\subsection{Extinction Correction, Absolute Peak Magnitudes and a Value for $H_0$} 
SN~1998bu exhibited an unusually high degree of reddening.  At
t$_{Bmax}$, $B-V=0.53\pm0.13$, much redder than the typical $B-V\approx0$
 of SNe~Ia (Branch, 1998). However, both the light curve shapes and
the spectral features of SN~1998bu are typical of a type~Ia event;
this and the results of our polarimetry lead us to conclude that
SN~1998bu was indeed a normal type~Ia supernova, but that it was
heavily reddened by dust.  Estimation of the amount of reddening is,
however, a difficult issue. Several methods have been considered.

One way is to use the relation between the equivalent width ($EW$) of
interstellar lines such as Na~I~D and the colour excess, $E(B-V)$
(Barbon {\it et al.} 1990).  Recently, Munari \& Zwitter (1997)
produced an improved determination of the relationship for the Milky
Way by measuring $E(B-V)$ for 32 O-type and B-type stars. The spectra
of SN~1998bu exhibit narrow Na~I~D absorption lines in the rest frames
of both the Milky Way and M~96 (Munari {\it et al.} 1998). The $EW$s
are, respectively, 0.19~\AA\ and 0.35~\AA. Centurion {\it et al.}
(1998) also reported {\it single} Na~I lines in the Milky Way and
M~96.  With these data, and assuming that Munari \& Zwitter (1997)
relation is also valid in M~96\footnote{Polarimetry results suggest
that the ISM in M~96 is similar to that of our own Galaxy, see
sub-section 2.3 } , we obtain a total colour excess ({\it i.e.}
including Galactic reddening) for SN~1998bu of $E(B-V)$=0.21, or
$A_{V}$=0.65 assuming $R_{V}$=3.1.  This is a significantly lower
value than that found by consideration of the SN colours (see below).
Moreover, Munari \& Zwitter (1997) warn that the use of the Na~I~D1
line to account for interstellar extinction is only valid when the
line can be modeled with a single Gaussian component.  When absorption
is multi-component, their relation provides an {\it upper limit} only
for $E(B-V)$. Thus, as also pointed out by Suntzeff {\it et al.}
(1999), this is inconsistent with the SN-colour derived values.

A different procedure is that followed by Phillips {\it et al.}
(1999). They use a technique based on the fact that all type~Ia events
show a very similar $B-V$ evolution between 30 and 90~days after
$V$-maximum.  For SN~1998bu, they find a total $E(B-V)=0.355\pm0.030$
or $A_{V}=1.10\pm0.09$ ($R_{V}$=3.1).  Another SN-colour procedure is
the Multicolor Light Curve Shape method, developed by Riess {\it et
al.} (1996). It is a multi-template ($BVRI$) method that uses a
training set of well-studied SNe~Ia to produce a ``standard SN~Ia'',
and deviations from this fiducial event are quantified as a function
of changes in luminosity and extinction. Application of this method by
Jha {\it et al.} (1999) to SN~1998bu gives a total $E(B-V)$=0.30 or
$A_{V}$=0.94, consistent with the value obtained by Phillips {\it et
al.}.  Yet another SN-colour method has been proposed by Krisciunas
{\it et al.} (2000).  They compared the $V-IR$ colours of SN~1998bu
with those of less-reddened SNe~Ia and infer $A_{V}=1.05\pm0.06$,
again consistent with Phillips {\it et al.}.

One further approach is to compare SN~1998bu with SN~1981B.  As
pointed out in Meikle \& Hernandez (2000) and in Section~3.1.1
(above), the detailed optical spectral features of these two SNe~Ia
around $t_{Bmax}$ are very similar, especially redward of 5000~\AA\
(see Figure~2).  The major difference between the two events is the
overall spectral slope.  We therefore assume that the Supernovae are
intrinsically identical and that the difference in slope is due to
extinction.  We modified a SN~1981B spectrum taken at t$_{Bmax}$
(Branch {\it et al.} 1983) by simultaneously scaling the flux by
wavelength-dependent and wavelength-independent factors.  The behavior
of the wavelength-dependent factor is taken to be the Cardelli {\it et
al.}  (1989) extinction law. This procedure was followed until a good
match was obtained between the SN~1981B spectrum and one of SN~1998bu
at -0.8~d. This was achieved with a {\it difference} in A$_V$ between
the two SNe~Ia of $\Delta A_V=0.60\pm0.06$, or $\Delta
E(B-V)=0.19\pm0.02$.  Clearly, this represents a lower limit to the
SN~1998bu extinction.  The matched spectra are shown in Fig.~9.  It
can be seen that there are some differences between the individual
spectral features but that they are generally quite small.  However,
as mentioned earlier, the calcium triplet absorption is significantly
deeper in SN~1981B.  Using their $B-V$ evolution method, Phillips {\it
et al.}  (1999) give a total $E(B-V)=0.13\pm0.03$ for SN~1981B.
Adding this to the extinction difference between SNe 1981B and 1998bu,
we obtain for SN~1998bu $E(B-V)=0.32\pm0.04$, or $A_V=1.0\pm0.11$, in
good agreement with the SN-colour derived values described above. We
used this value with the Cardelli {\it et al.} (1989) law, $R_V=3.1$,
and a distance modulus of $30.25\pm0.18$ (see Introduction) to
determine the absolute intrinsic peak magnitudes.  These are shown in
Table~6.

We find that SN~1998bu peaked at $V=-19.37\pm0.23$ (see Table~6 for
other bands).  This is $0.26\pm0.30$ and $0.05\pm0.32$ fainter than
the values found by, respectively, Suntzeff {\it et al.} and Jha {\it
et al.}  Applying our values for the absolute peak $BVI$ magnitudes
and reddening-corrected $\Delta$$m_{15}$(B) to Phillips {\it et al.}'s
(1999) relations (17--19) we obtain $H_0=70.4\pm4.6$~km/s/Mpc.
Suntzeff {\it et al.} derive $H_0=64\pm2.2(int.)\pm3.5(ext.)$~km/s/Mpc
from their SN~1998bu light curves.  The barely significant difference
between our value and that of Suntzeff {\it et al.} can be explained
by the choices of extinction correction and distance modulus.

\section{Summary}
We have presented first-season $UBVRIJHK$ photometry and $UBVRI$
polarimetry of the nearby type~Ia SN~1998bu.  Also presented are a set
of optical spectra spanning ~$t=-6.8~d$ to $t=+19.2d$ plus a single
IR spectrum at ~$t=+35.5~d$. The optical light curve shapes are typical
of a normal type~Ia supernova.  In addition, de-reddening of the
-0.8~day optical spectrum using the standard Galactic extinction law
(Cardelli {\it et al.}  1989) produces a spectrum which is highly
similar to those of classic SNe~Ia.  This suggests strongly that the
very red colour of SN~1998bu is due to extinction. It is also likely
that the relatively strong polarization is associated with grains
within M~96 along the line of sight.  We conclude that both the light
curve shapes and spectra indicate that SN~1998bu is a {\it normal}
type~Ia supernova.

The $BVRI$ peak magnitudes we obtained are consistent with those of
Suntzeff {\it et al.} (1999) and Jha {\it et al.} (1999).  We
de-reddened the photometry using the Cardelli {\it et al.} law with
$R_V=3.1$ and A$_{V}$ derived from estimates of intrinsic SN~Ia
colours.  The absolute peak magnitudes were then found using a
distance modulus of $30.25\pm0.18$.  We find that SN~1998bu peaked at
$V=-19.37\pm0.23$ (see Table~6 for other bands). Our results yield a
value for the Hubble Constant of $H_0=70.4\pm4.3$~km/s/Mpc.

Combination of our IR photometry with those of Jha {\it et al.}
provides one of the most complete early-phase IR light curves for a
SN~Ia published so far.  In particular, SN~1998bu is the first normal
SN~Ia for which good pre-t$_{Bmax}$ IR coverage has been obtained.  It
reveals that the $JHK$ light curves peak about 5~days earlier than in
the $B$-band.  Secondary maxima are seen in the $IJHK$ bands, with a
corresponding inflection in the $R$-band. For further details see
Meikle (2000).

\subsection*{Acknowledgements}

We are indebted to all those observers who contributed to this work by
giving up some of their telescope time. We are especially grateful to
NOAO WIYN Queue observers P. Smith, D. Willmarth and D. Harmer. Our
gratitude also goes to R. Barrena-Delgado for his observations at
IAC80. We thank D. Branch and N. Tanvir for helpful comments.  MH is
supported by PPARC (UK) and TMR (EU).  MRB is supported by PPARC.
This work was in part supported by the Portuguese Foundation for
Science and Technology (grant PESO/P/PESO/1196/97).

The data provided by EXPORT was obtained during the 1998 International
Time of the Canary Islands Observatories awarded to EXPORT, with the
observations and data reduction being carried out by EXPORT members
D. de Winter, F. Garzon, L. F. Miranda and R. Oudmaijer.

The IAC80 and TCS are operated on the island of Tenerife by the
Spanish Observatorio del Teide of the Instituto de Astrofisica de
Canarias.

UKIRT is operated by the Joint Astronomy Centre on behalf of the
U.K. Particle Physics and Astronomy Research Council.

The William Herschel, Isaac Newton and Jacobus Kapteyn telescopes are
operated on the island of La Palma by the Isaac Newton Group in the
Spanish Observatorio del Roque de los Muchachos of the Instituto de
Astrofisica de Canarias.

The WIYN Observatory is a joint facility of the University of
Wisconsin-Madison , Indiana University, Yale University, and the
National Optical Astronomy Observatories.

\newpage

{\bf FIGURE CAPTIONS}

{\bf Figure 1:} SN~1998bu in M~96 in the $R$-band, 1998 June 8 (WIYN). The
five comparison stars (CS) are labelled. The field of view is
$6.5\times6.5$ arcmin.  North is up and East is to the left. 
\newline
\newline
{\bf Figure 2:} Optical spectra of SN~1998bu taken at the WHT and INT (La Palma) and
the WIYN Telescope (Kitt Peak) (see Table~5 for details). The spectra
have not been corrected for redshift or reddening. The epochs are with
respect to t$_{Bmax}$ = 1998 May~19.8 = 0~days.  For clarity, the
spectra have been displaced vertically. The dotted lines on the left
side indicate zero flux for each of the spectra. For +19.2~d zero flux
is at the x-axis.  The lowest dotted line indicates zero flux for both
the +11.5~d and +13.5~d spectra.  In general, the absolute fluxing is
accurate to $\pm$10\%. For --6.8~d the uncertainty is closer to
$\pm$15\%.
\newline
\newline
{\bf Figure 3:} Infrared spectrum of SN~1998bu taken at UKIRT at
+35.5~days.  The epoch is with respect to t$_{Bmax}$ = 1998
May~19.8~UT.  The data have not been corrected for redshift or
reddening.
\newline
\newline
{\bf Figure 4:} Illustration of the high degree of similarity in the
optical spectra of the type~Ia Supernovae SNe~1981B, 1994D and 1998bu.
The epochs are all within about 1~day of t$_{Bmax}$.  To aid the
comparison the spectra have been scaled and shifted vertically by
arbitrary amounts and have been wavelength shifted to the local
standard of rest for the respective supernovae. Their zero flux axes
are indicated by the dotted lines on the left hand axis.  The SN~1981B
spectrum is from Branch {\it et al.} (1983) and is courtesy of
B.~Leibundgut and P.~Nugent.  The SN~1994D spectrum is from Meikle
{\it et al.} (1996).
\newline
\newline
{\bf Figure 5:} Infrared spectrum of SN~1998bu compared with those of
other type~Ia Supernovae at a range of epochs (Bowers {\it et al.},
1997).  Also shown is the IR spectrum of SN~1998bu at $\sim$25~d
published by Jha {\it et al.} (1999).  The spectra have been shifted
vertically and scaled for clarity.  Their zero flux axes are indicated
by the dotted lines on the left hand axis.  The spectra have been
wavelength shifted to the local standard of rest for the respective
Supernovae.
\newline
\newline
{\bf Figure 6:} As Figure~5 but expanded to reveal the detail in the spectra 
longward of 11,000\AA.
\newline
\newline
{\bf Figure 7:} Optical light curves for SN~1998bu.  For clarity they have
been vertically displaced by the amounts indicated.  The epoch of
maximum blue light, t$_{Bmax}$, corresponds to 1998~May~19.8~UT.
\newline
\newline
{\bf Figure 8:}Infrared and optical light curves for SN~1998bu.  For
clarity they have been displaced vertically by arbitrary amounts.  The
IR photometry was obtained at the OAN (Mayya \& Puerari (1998)), TCS,
IRTF, UKIRT and WHT telescopes.  The optical light curves are as
plotted in Figure~7.  Also shown are template light curves in $BV$
(Leibundgut 1988), $RI$ (Schlegel 1995) and $JHK$ (Elias {\it et al.}
1985). The $BVRI$ templates were shifted in both axes to give the best
match to the data.  The $JHK$ templates were shifted only vertically.
Their horizontal position was fixed by the epoch of t$_{Bmax}$ as
indicated in Elias {\it et al.}  (see text).
\newline
\newline
{\bf Figure 9:}  This illustrates the determination of the relative
extinction to SNe~1981B and 1998bu.  The solid and dotted lines show
the optical spectra of SNe 1998bu and 1981B respectively.  The dashed
line shows the SN~1981B spectrum reddened by A$_{V}=0.60\pm0.07$ to
match that of SN~1998bu, using the extinction law of Cardelli {\it et
al.} (1989), with $R_V=3.1$.

\bsp

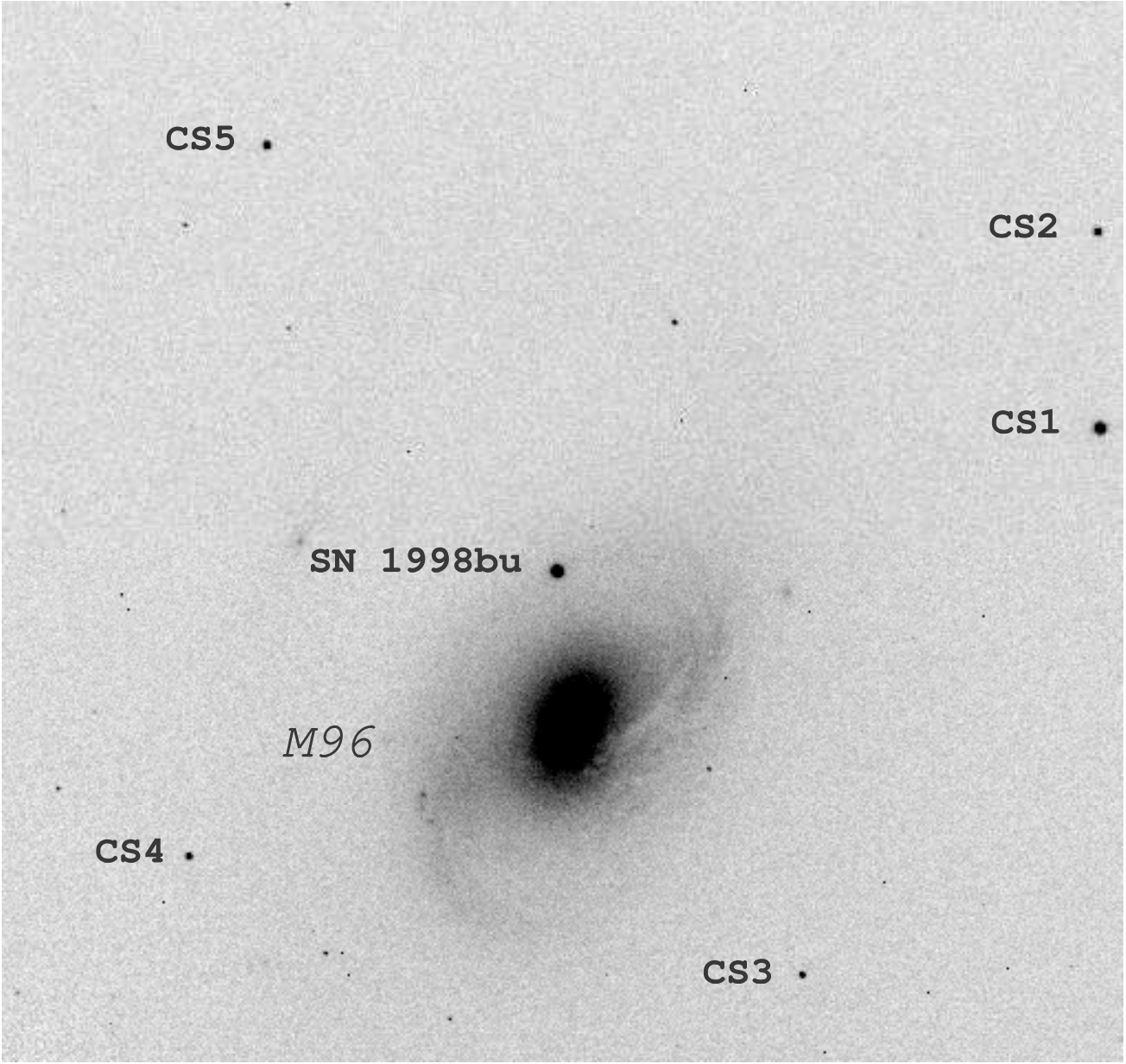
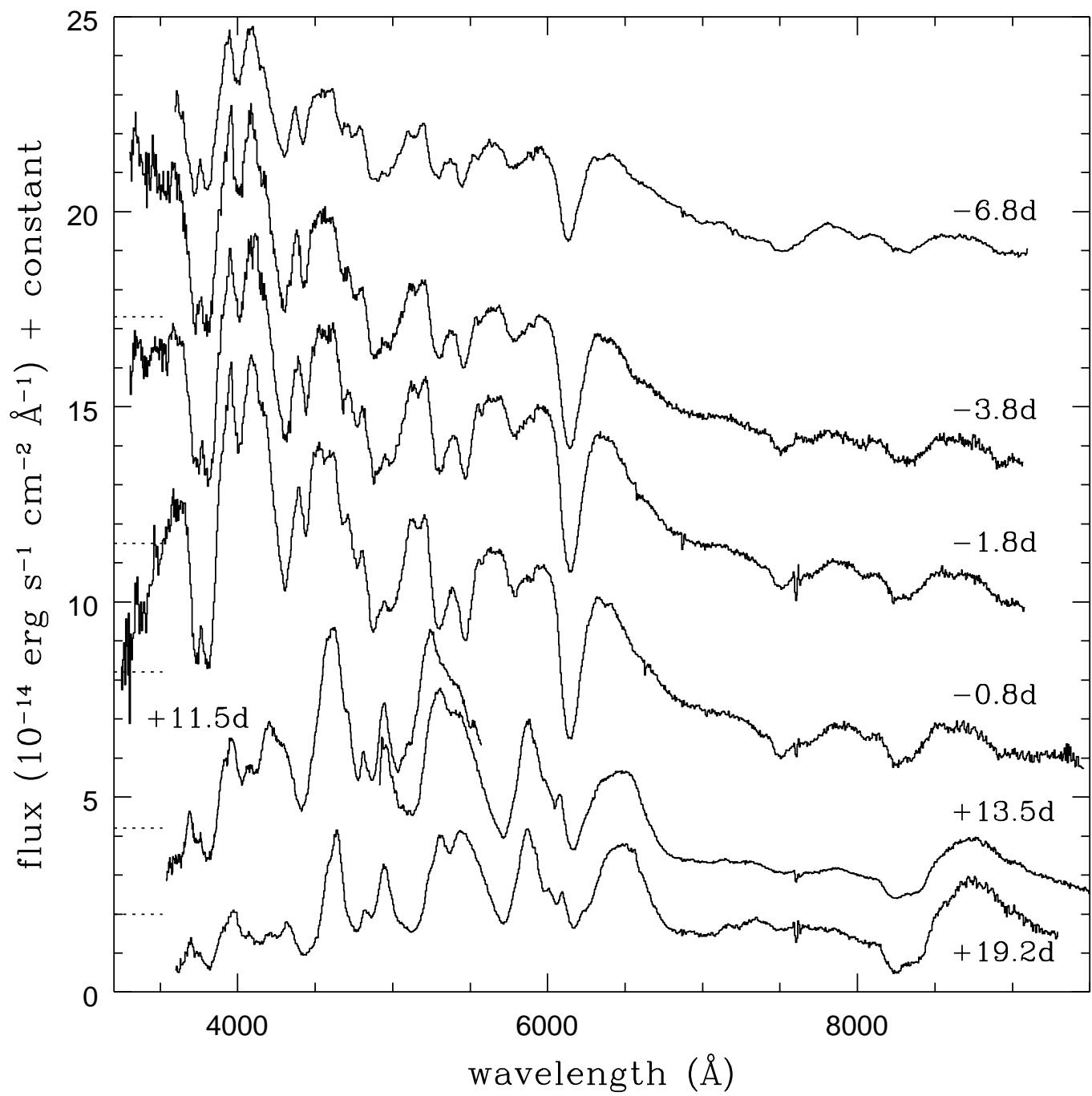
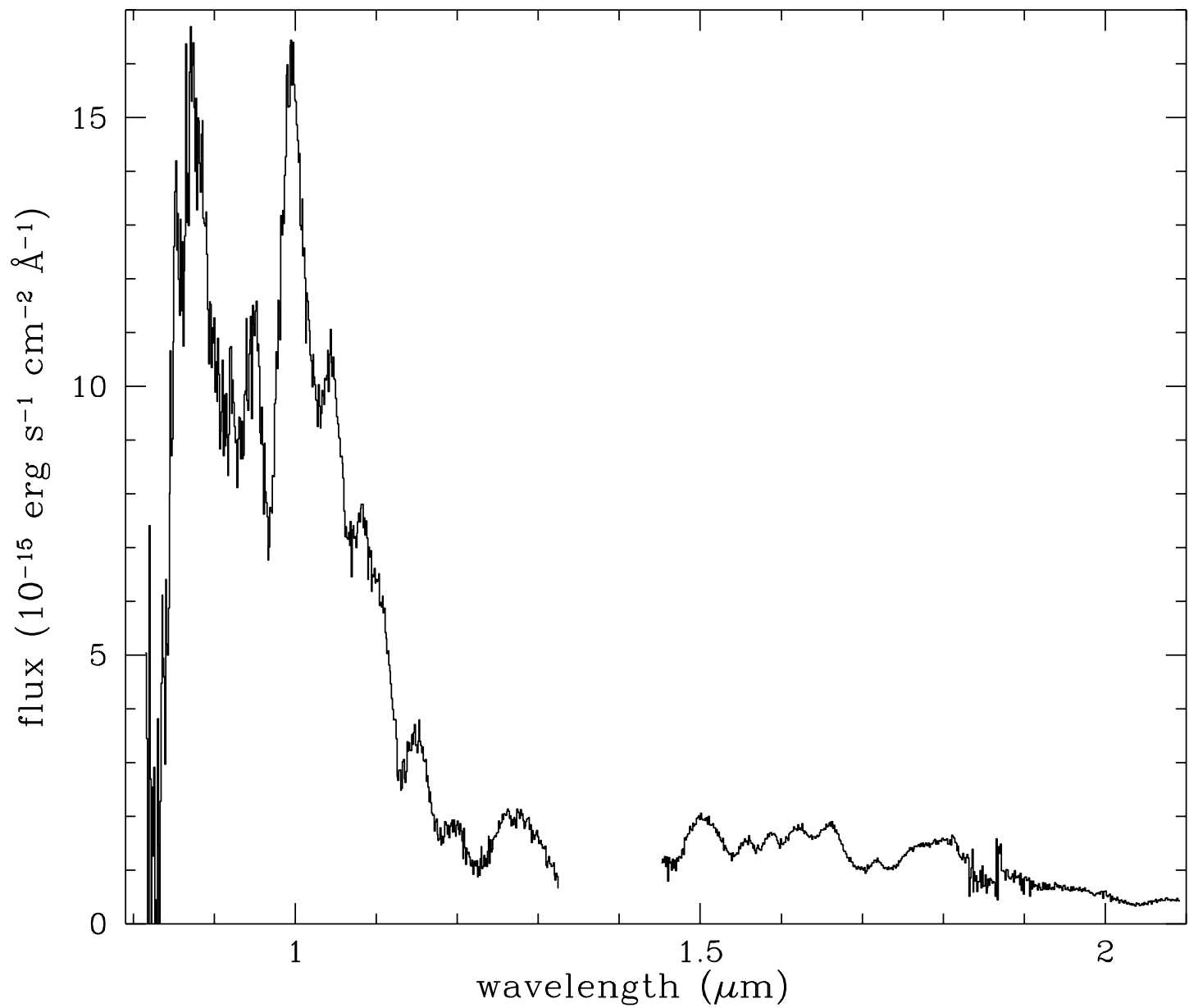
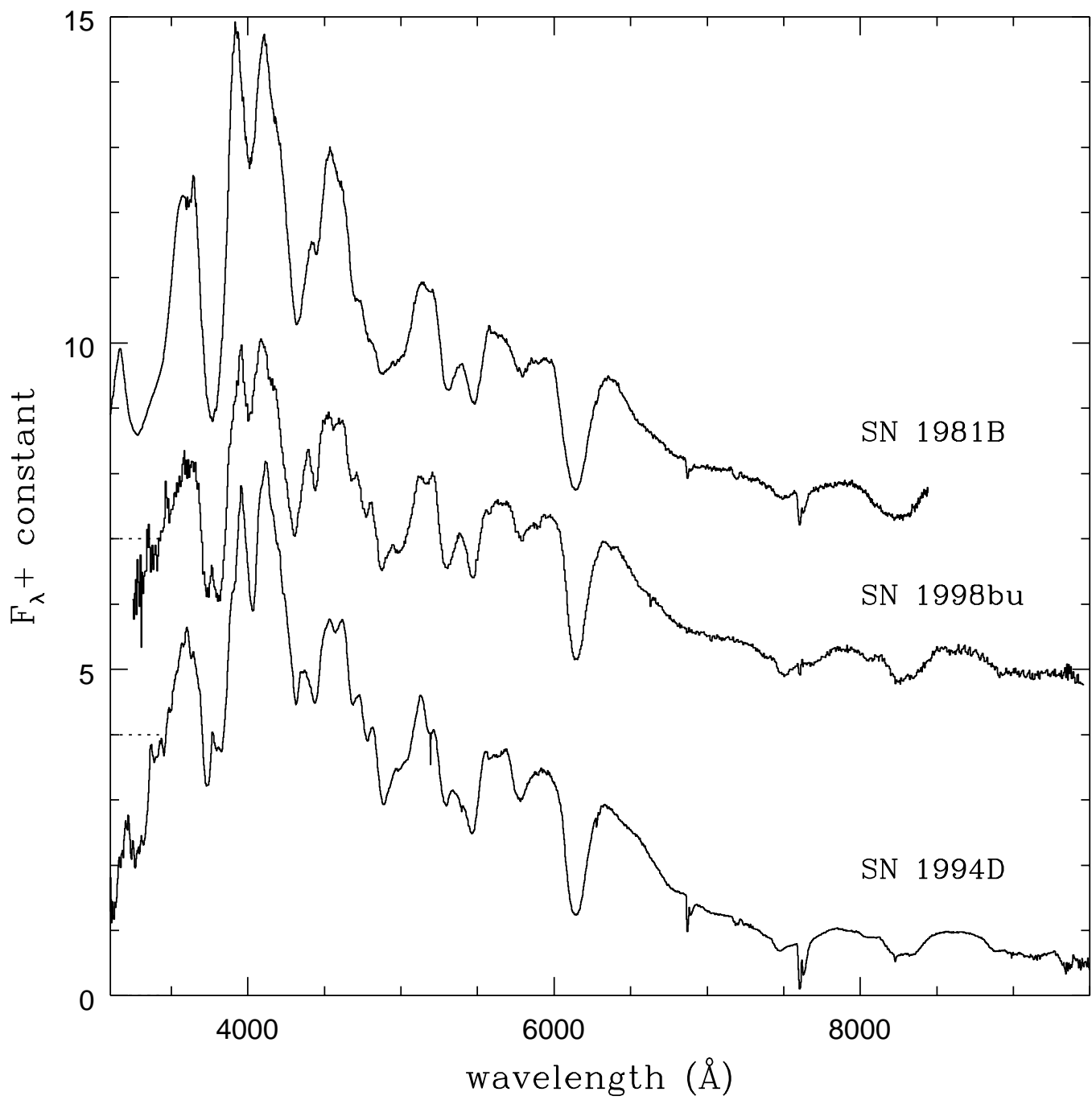
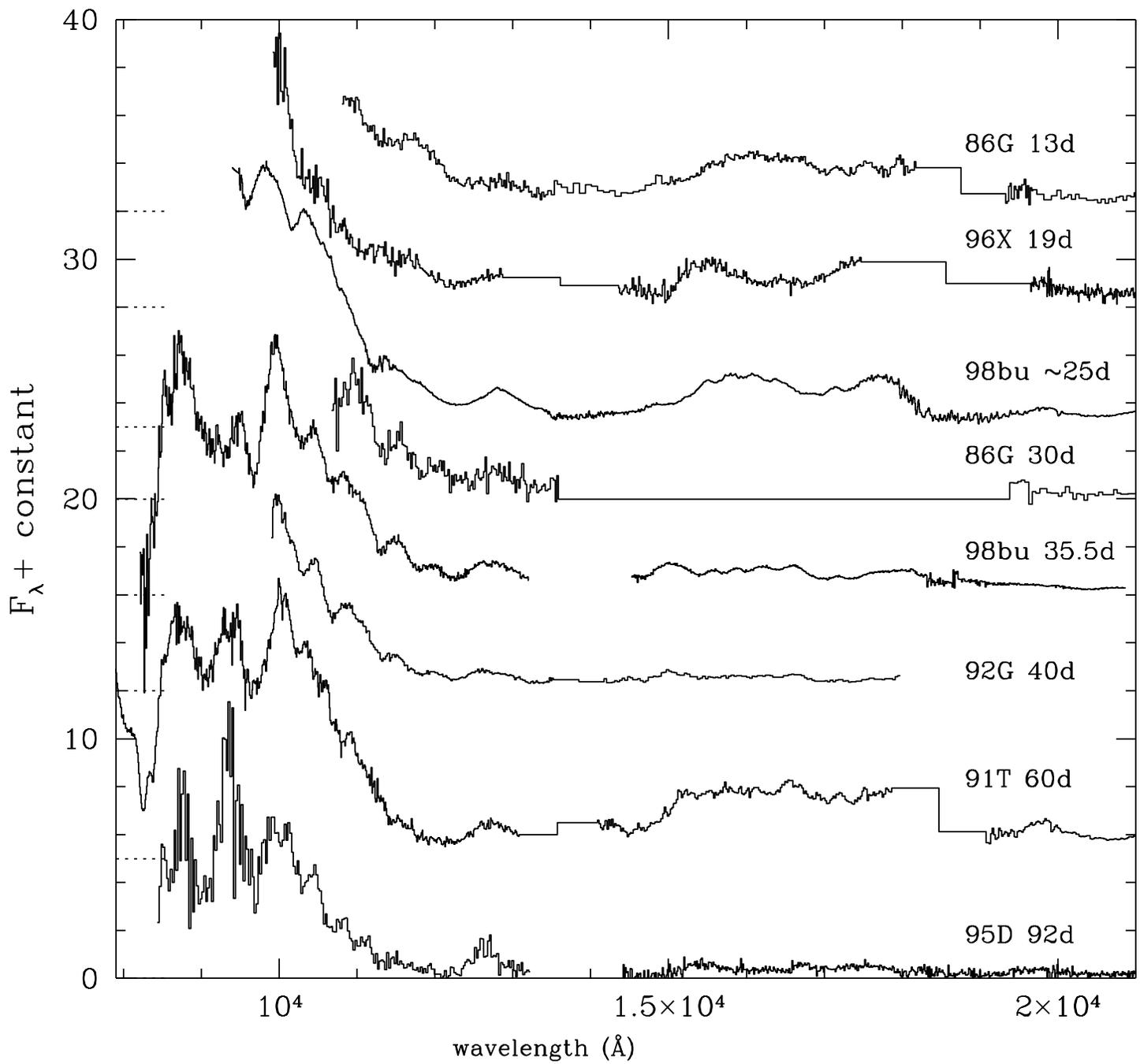
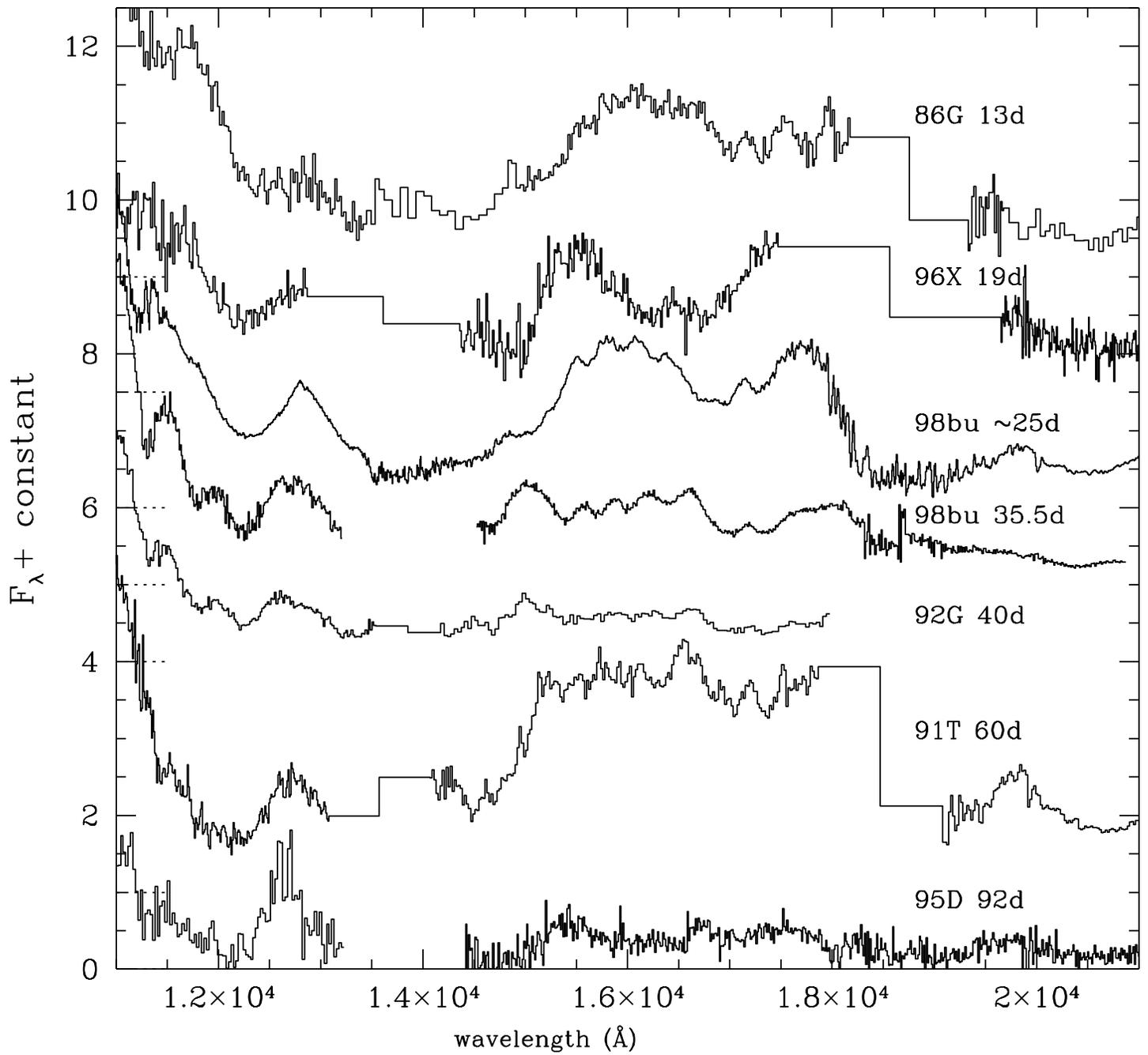
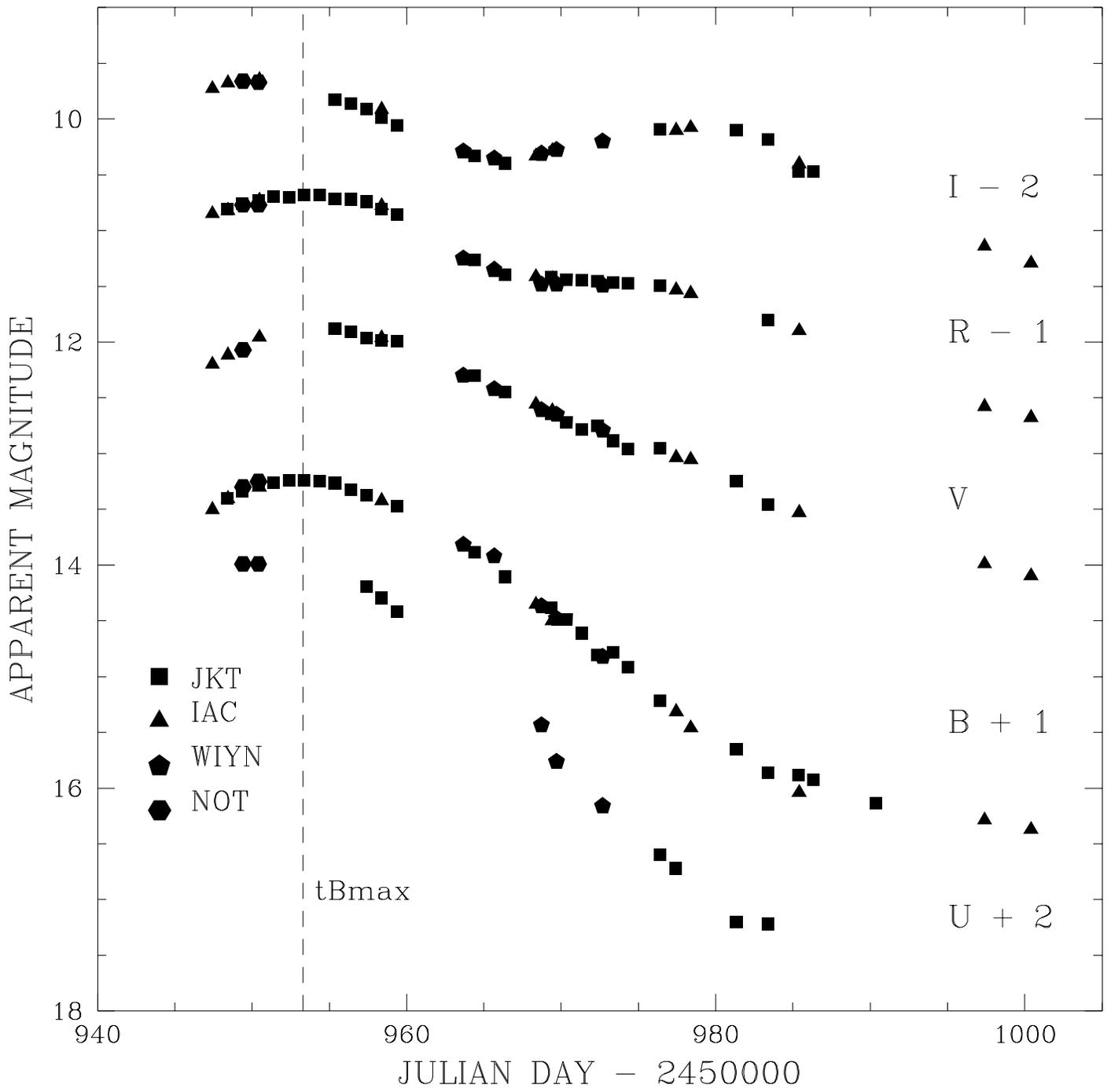
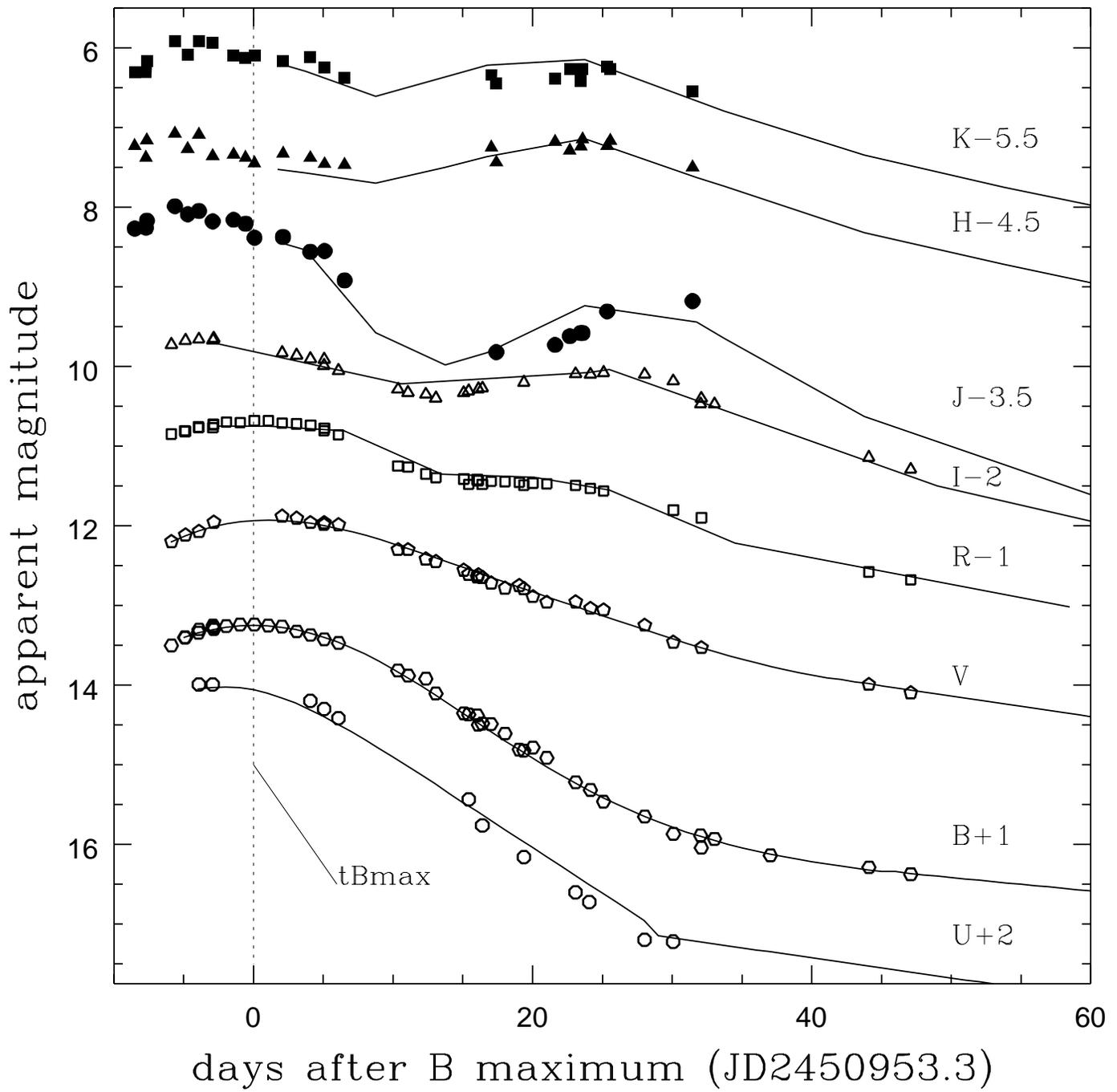
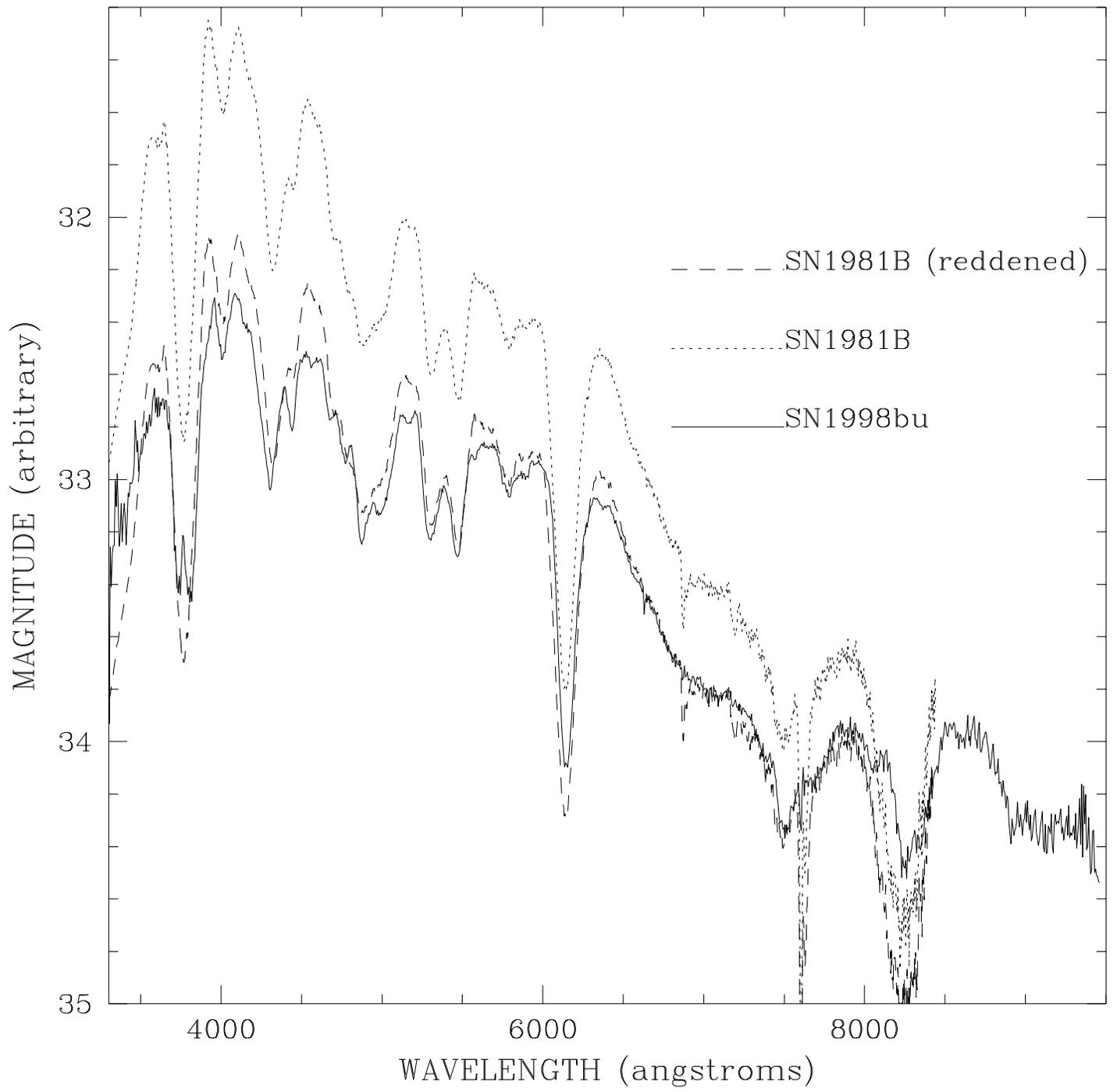
\end{document}